\documentclass[
 reprint,
longbibliography,
 amsmath,amssymb,
 aps,
 pra,
]{revtex4-1}

\usepackage{mathtools}
\usepackage{graphicx}
\usepackage{dcolumn}
\usepackage{bm}
\usepackage{dsfont}
\usepackage{rotating}
\usepackage[table]{xcolor}
\usepackage{epstopdf}
\usepackage{array}
\usepackage{booktabs}
\usepackage{colortbl}
\newcolumntype{C}{>{$}c<{$}}
\usepackage[%
  colorlinks=true,
  urlcolor=blue,
  linkcolor=blue,
  citecolor=blue
]{hyperref}
\usepackage{soul}

\newcommand{\tr}[1]{\mathrm{tr}\left(#1\right)}
\newcommand{\tra}[1]{\mathrm{tr}_{A}\left(#1\right)}
\newcommand{\trd}[1]{\mathrm{tr}\left[#1\right]}
\newcommand{\trad}[1]{\mathrm{tr}_{A}\left[#1\right]}
\newcommand{\id}{\mathds{1}}
\newcommand{\ie}{\textit{i.e.}}
\renewcommand{\b}{\mathsf{b}}
\newcommand{\y}{\mathsf{y}}
\newcommand{\p}{\mathsf{p}}
\newcommand{\corr}[1]{\left< #1 \right>}
\newcommand{\R}{\mathds{R}}
\newcommand{\C}{\mathds{C}}
\newcommand{\de}[1]{\left( #1 \right)}
\newcommand{\ket}[1]{\left| #1 \right>}
\newcommand{\bra}[1]{\left< #1 \right|}
\newcommand{\proj}[1]{\ket{#1}\!\!\bra{#1}}
\newcommand{\eg}{\textit{e.g.}}
\newcommand{\DE}[1]{\left\{ #1 \right\}}
\newcommand{\Hil}{\mathcal{H}}

\begin{document}

\preprint{APS/123-QED}

\title{Measurement compatibility in Bell nonlocality tests}

\author{Tassius Temistocles}
\affiliation{Departamento de F\'{i}sica, Instituto de Ci\^{e}ncias Exatas, Universidade Federal de Minas Gerais, Caixa Postal 702, CEP 30123-970, Belo Horizonte, Brazil}
\affiliation{Instituto de Matem\'{a}tica, Estat\'{i}stica e Computa\c{c}\~{a}o Cient\'{i}fica, Universidade Estadual de Campinas, CEP 13083-859, Campinas, Brazil}

\author{Rafael Rabelo}
\affiliation{Instituto de F\'{i}sica ``Gleb Wataghin'', Universidade Estadual de Campinas, CEP 13083-859, Campinas, Brazil}

\author{Marcelo {Terra Cunha}}
\email{tcunha@ime.unicamp.br}
\affiliation{Instituto de Matem\'{a}tica, Estat\'{i}stica e Computa\c{c}\~{a}o Cient\'{i}fica, Universidade Estadual de Campinas, CEP 13083-859, Campinas, Brazil}%

\date{\today}

\begin{abstract}
{Incompatibility} of observables, or measurements, is one of the key features of quantum mechanics, related, among other {concepts}, to Heisenberg's uncertainty relations and Bell nonlocality. In this manuscript we show, however, that even though {incompatible} measurements are necessary for the violation of any Bell inequality, some relevant Bell-like inequalities may be obtained if compatibility relations are assumed between the local measurements of one (or more) of the parties. Hence, compatibility of measurements is not necessarily a drawback and may, however, be useful for the detection of Bell nonlocality {and device-independent certification of entanglement}.
\end{abstract}

\maketitle

\section{Introduction}
Quantum theory is fundamentally distinct from any classical theory of physics, a fact that is well known and accepted nowadays. Even though there is no consensus regarding a physical principle that explains such departure, the distinction between quantum and classical mechanics is clear at the level of their mathematical formalisms. For instance, two of the most interesting nonclassical features present in quantum theory are entanglement and incompatibility between measurements.

The fact that there are measurements, or observables, that are incompatible, {which can not be jointly treated as one observable within quantum theory}, is one of the key ingredients behind some of the most astonishing phenomena related to nonclassicality, such as Bell nonlocality \cite{Bell1964,Brunner2014} and Bell-Kochen-Specker contextuality \cite{Bell1966, Kochen1966}. Both arose from investigations regarding the completeness of quantum theory \cite{Einstein1935} and are related to stronger-than-classical correlations between outcomes of measurements performed on quantum systems. {The theory of Bell nonlocality and the theory of Bell-Kochen-Specker contextuality}, though, have been developed independently, in a sense, and each presents its own particular features. In this manuscript we focus our attention on the former, although the reader may notice that there will be elements of the latter.

The paradigmatic example of Bell nonlocality of quantum systems takes place in a bipartite measurement scenario, where two characters, Alice and Bob, are able to {choose between} two possible dichotomic measurements {to perform} on their respective share of a previously prepared joint system. Denoting by $A_{x} \in \{\pm1\}$ the outcome of measurement $x \in \{0,1\}$ of Alice, and by $B_{y} \in \{\pm1\}$ the outcome of measurement $y \in \{0,1\}$ of Bob, any so-called \emph{local hidden variable} (LHV) theory of joint probabilities that govern the behavior of the measurement devices will lead to mean values that necessarily obey the \emph{Clauser-Horne-Shimony-Holt} (CHSH) \cite{Clauser1969} inequality
\begin{equation}
\corr{A_{0}(B_{0}+B_{1}) + A_{1}(B_{0}-B_{1})} \leq 2.
\end{equation}
If the measurements are performed on quantum systems, however, then the inequality does not need to be respected, and a violation of up to the value of $2\sqrt{2}$ may be observed \cite{Tsirelson1980}. This shows that quantum theory is incompatible with LHV theories, an observation first made by Bell \cite{Bell1964}.

Two necessary conditions for Bell nonlocality to manifest in quantum systems are (i) entanglement, in the state of the shared system; and (ii) incompatibility between the measurements, in each party. Curiously, neither (i) \cite{Werner1989} nor (ii) \cite{Hirsch2018} is a sufficient condition. Regarding (i), it became an important question in the field (for both fundamental and practical reasons) to identify which entangled states could ultimately lead to Bell nonlocality. On one hand, there has been an effort to obtain examples of local entangled states, those that never lead to nonlocal correlations \cite{Barrett2002,Almeida2007,Cavalcanti2016,Hirsch2016,Oszmaniec2017}. On the other hand, several nonstandard measurement scenarios have been proposed where even local entangled states can lead to Bell nonlocality, and concepts like hidden-nonlocality \cite{Popescu1995, Peres1996, Liang2012, Gallego2014, Hirsch2013} and activation of nonlocality \cite{Cavalcanti2011, Cavalcanti2012, Palazuelos2012, Cavalcanti2013} have been defined.

In this manuscript we propose a measurement scenario where we explicitly assume the existence {of subsets of compatible measurements {for} (at least) one of the parties}. We show how this assumption drastically changes the measurement scenario and leads to a range of Bell-like inequalities that are potential tools to improve on condition (i), leading to examples of entangled states that have not been known to be nonlocal. We present an example, in one of the simplest scenarios of this approach, where, without the compatibility assumption, the only relevant Bell inequality is CHSH \cite{Pironio2014}, and, with the compatibility assumption, 26 Bell-like inequalities arise.
{We explicitly show that one of these inequalities reveals Bell nonlocality in {two families} of quantum states for regions of parameters where {the CHSH inequality is not violated. Also, there is numerical evidence that, in part of this region, the states {of one of such families} do not violate the I$_{3322}$ inequality \cite{i3322} either.}

{In addition to revealing the nonlocality of quantum states that would not be displayed in standard Bell scenarios, multipartite measurement scenarios with local compatible measurements are interesting for both fundamental and practical reasons: fundamentally, because these are the scenarios that are suitable for the joint study and observation of Bell nonlocality and Bell-Kochen-Specker contextuality; practically, because, in some particular cases (as in the example we present), the compatibility relations can be implemented in a device-independent manner, and so such scenarios may be useful for the implementation of stronger device-independent information processing protocols, such as device-independent certification of entanglement, as in the example we present.}


\section{The scenario}\label{sec:scenario}
Consider a scenario where two parties, Alice and Bob, are able to perform different measurements on their subsystems of a shared physical system. The measurements are implemented by black-boxes, of which only classical inputs and outputs are available to the users. The inputs and outputs of Alice's box are labeled $x \in \mathcal{X}$ and $a \in \mathcal{A}$, respectively, and the inputs and outputs of Bob's box are labeled $y \in \mathcal{Y}$ and $b \in \mathcal{B}$, respectively. {Since the users do not have access to the inner workings of the measurement devices, but only to their classical inputs and outputs, the best description of the experiment is given by the} joint probabilities $p(a,b|x,y)$ of the parties to observe outputs $a$ and $b$, on the condition that inputs $x$ and $y$ are chosen, respectively. The collection of probabilities $ p(a,b|x,y) $ for all $a\in \mathcal{A}$, $b \in \mathcal{B}$, $x\in \mathcal{X}$, $y \in \mathcal{Y}$ is referred as the \emph{behavior} {or the \emph{empirical model}} of the box, and will be denoted $\p$.

A behavior is said to be \emph{no-signalling} if the choice of input of one of the parties cannot influence the marginal probability distribution of outcomes of the other, \ie, it satisfies the following \emph{no-signalling conditions}:
\begin{subequations}
\begin{align}
p(a|x) = & \sum_{b}p(a,b|x,y) = \sum_{b}p(a,b|x,y^{\prime}),\\
p(b|y) = & \sum_{a}p(a,b|x,y) = \sum_{a}p(a,b|x^{\prime},y).
\end{align}
\end{subequations}
More restrictedly, a behavior is said to be \emph{local} if there exist a variable $\lambda$, a probability distribution $q(\lambda)$, and probability distributions $p(a|x,\lambda)$ and $p(b|y,\lambda)$ such that
\begin{equation}\label{eq:local}
p(a,b|x,y) = \sum_{\lambda} q(\lambda) p(a|x,\lambda) p(b|y,\lambda).
\end{equation}
It can be shown that the probability distributions $p(a|x,\lambda)$ and $p(b|y,\lambda)$ can be made deterministic without loss of generality.

If the boxes perform measurements on quantum systems, then the probabilities are given by Born's rule:
\begin{equation}\label{eq:Born}
p(a,b|x,y) = \tr{\rho P_{a|x}\otimes Q_{b|y}},
\end{equation}
where $\rho$ is the density operator that describes the state of the joint system {while} $P_{a|x}$ and $Q_{b|y}$ are, in general, POVM effects. It is well known that, in Bell scenarios, the set of local behaviors is strictly contained in the set of quantum behaviors, which, in its turn, is strictly contained in the set of no-signalling behaviors. 

Now, consider the single black box of Bob, with inputs $y \in \mathcal{Y}$, and outputs $b \in \mathcal{B}$. 
Suppose, however, that some measurements are \emph{compatible}, and that each set of compatible measurements {defines} a \emph{context}, $\y \subset \mathcal{Y}$ -- {\textit{sans serif} types refer to labels that represent ordered tuples of the corresponding \textit{serif} labels, \textit{e. g.}, $\y = (y_{i},\dots,y_{j})$, $\b = (b_{k},\dots,b_{l})$.} Let $\mathcal{C} = \{\y\}$ denote the set of possible contexts of the scenario; {it is usual to represent this set by means of the \textit{compatibility hypergraph} $G = (V,E)$, where each measurement is associated to a vertex $v \in V$ and each context associated to a hyperedge $e \in E$}. In particular, in scenarios where the contexts have cardinality $2$, $G$ will take the form of a regular graph. Compatible measurements can be jointly performed, and a joint probability distribution of the outcomes can be defined. Let $\b$ denote the ordered outcomes of the measurements in a context $\y$. Then, the behavior of this single box is best described by the probabilities $p(\b|\y)$. Noting that an individual measurement can appear in more than one context, it is usual to assume that the marginal behavior of each individual measurement $y \in \mathcal{Y}$ to be well defined, regardless of the context. This leads to the so-called \emph{no-disturbance} conditions:
\begin{equation}\label{eq:ND}
p(b|y) = \sum_{{\b/b}} p(\mathsf{b}|\mathsf{y}) = \sum_{{\b/b}} p(\mathsf{b}|\mathsf{y}^{\prime}),
\end{equation}
{where $\b/b$ denotes all labels in $\b$ except $b$}, for all $b$ in $\mathcal{B}$, for all $y \in \mathcal{Y}$, and for all $\mathsf{y}, \mathsf{y}^{\prime} \in \mathcal{C}$ such that {$y \in \mathsf{y} \cap \mathsf{y}^{\prime}$}.

Suppose, now, a bipartite scenario, as considered previously, but let Bob be able to perform joint measurements according to given compatibility rules that lead to a set of contexts $\mathcal{C}$. We refer to this scenario as a \emph{Bell scenario extended with compatible measurements}, or \emph{extended Bell scenario}, for short. Then, the joint behavior of the boxes will be given by probabilities $p(a,\b|x,\y)$, for all $a \in \mathcal{A}$, $\b \in \mathcal{B}^{|\y|}$, $x \in \mathcal{X}$ and $\y \in \mathcal{C}$. We assume the behavior to be no-signalling, and the marginal, local behavior of Bob's box to obey the no-disturbance conditions. This measurement scenario is illustrated in the lower panel of Fig. \ref{fig:new_scenario}.

\begin{figure}[htbp]
\includegraphics[width = 0.8\columnwidth]{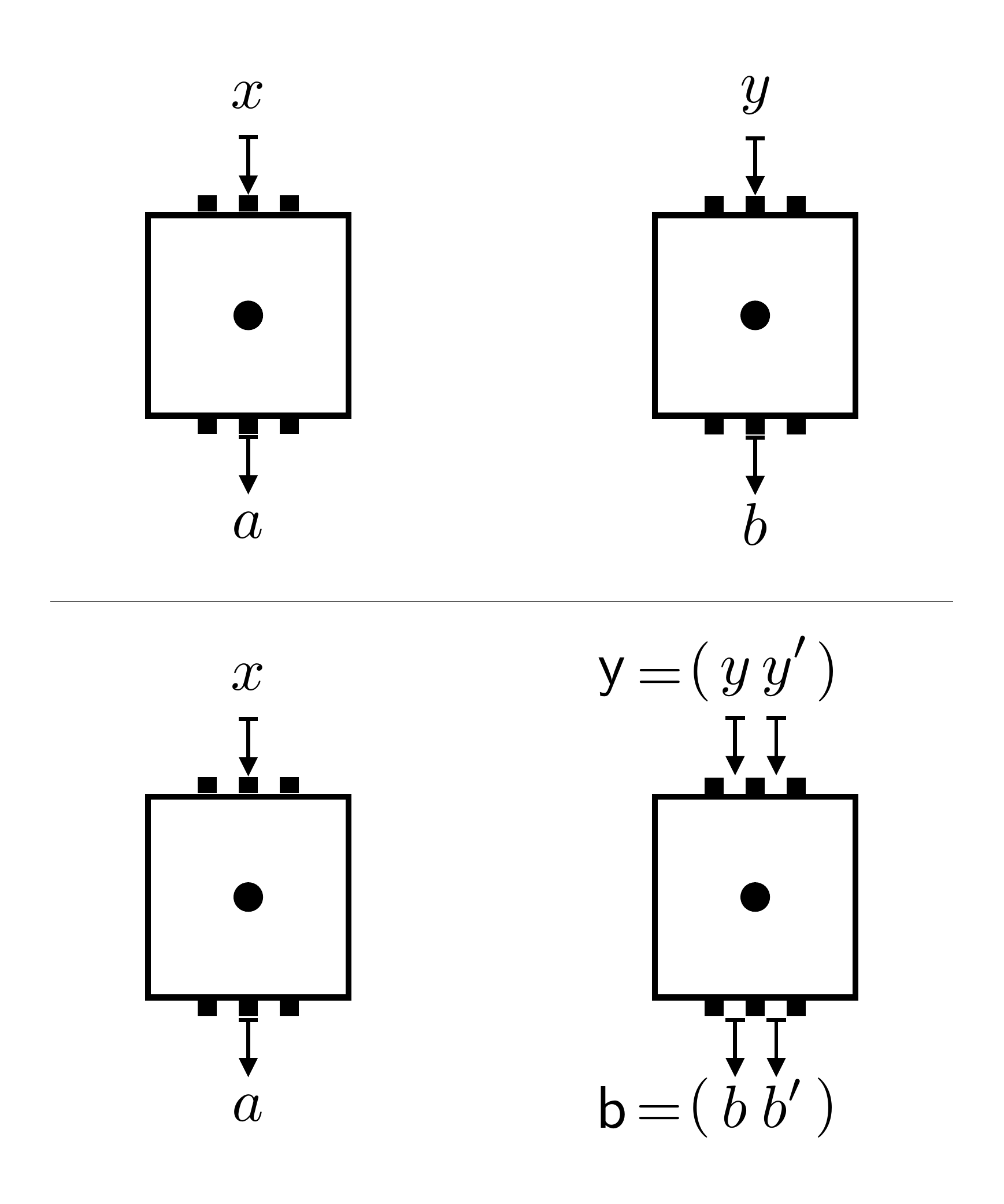}
\caption{Upper panel: a standard Bell nonlocality scenario. In each round of the experiment, party A (B) chooses measurement $x$ ($y$) to perform on its respective subsystem, obtaining outcome $a$ ($b$). The experiment is described by the conditional probabilities $p(a,b|x,y)$. Lower panel: a Bell nonlocality scenario where party B is able to perform compatible measurements. Now, in each round, party B chooses two (or more, according to the context) measurements to be jointly performed, $y$ and $y'$, obtaining outcomes $b$ and $b'$, respectively. The experiment is described by the conditional probabilities $p(a,b,b'|x,y,y')$; defining $\y = (y,y')$ and $\b = (b,b')$, the same conditional probabilities can be written as $p(a,\b|x,\y)$.}
\label{fig:new_scenario}
\end{figure}

We define a behavior to be local in this scenario if there are a variable $\lambda$, a probability distribution $q(\lambda)$, and probability distributions $p(a|x,\lambda)$ and $p(\b|\y, \lambda)$ such that
\begin{equation}\label{eq:new_local}
p(a,\b|x,\y) = \sum_{\lambda} q(\lambda) p(a|x,\lambda) p(\b|\y,\lambda).
\end{equation}
Here, $p(a|x,\lambda)$ can be assumed to be deterministic probability distributions. This is due to the fact that the set of marginal behaviors $\p_{A} = \{p(a|x)\}$ is a convex set with finitely many extremal points, \ie, a \emph{polytope}, whose vertices are exactly the deterministic distributions that suffice in the definition; let $\mathcal{P}_{A}$ denote this polytope. The set of marginal behaviors $p_{B} = \{p(\b|\y)\}$ is also a polytope, {since it is characterized by the intersection of a finite number of subspaces, given by the no-disturbance conditions Eq. \eqref{eq:ND}, together with nonnegativity and normalization conditions of probability distributions. As a polytope, it has a finite number of extremal points; however, they are not necessarily deterministic probability distributions}. Let $\mathcal{P}_{B}$ denote this polytope, the \emph{no-disturbance polytope} of party $B$. It is easy to see, then, that, in this definition, the joint behavior $\p = \{p(a,\b|x,\y)\}$ will be a convex combination of a finite set of points, so the set of $\p$ will also be a polytope, $\mathcal{P}_{AB}$, {the \emph{local and no-disturbance  polytope}}; its vertices are all possible ``products'' between the vertices of $\mathcal{P}_{A}$ and $\mathcal{P}_{B}$.

Now, given that the set of local behaviors is a polytope whose vertices are known (provided the vertices of $\mathcal{P}_{B}$ are known), we can change its representation by means of specialized software, such as \texttt{porta} \cite{porta} or \texttt{panda} \cite{panda}, and obtain the inequalities associated to its facets. Such inequalities will be Bell-like inequalities whose violation certify Bell nonlocality{, in the sense that these correlations can not be explained locally, by Eq.~\eqref{eq:new_local}}.


\section{Compatible measurements on quantum systems}
In the measurement scenarios introduced in the previous section, it is assumed that compatible measurements are performed by one or more of the parties. Thus, quantum realizations of such measurement scenarios require quantum measurements that are compatible according to the predefined contexts.

In quantum theory, compatibility of measurements is usually defined in terms of \emph{observables}. An observable is an Hermitian operator acting on the Hilbert space of the quantum system, that is related to a projective measurement by means of its spectral decomposition. Two observables $B_{y}$ and $B_{y'}$ are said to be compatible if they commute, $\left[ B_{y}, B_{y'} \right] = B_{y}B_{y'} - B_{y'}B_{y} = 0$. This condition implies that both operators can be diagonalized in the same basis, and that a third operator that represents the joint action of them can be defined.

Let $B_{y} = \sum_{b}b Q_{b|y}$ and $B_{y'} = \sum_{b'}b'R_{b'|y'}$ be the spectral decompositions of observables $B_{y}$ and $B_{y'}$, {where $b$ and $b'$ are their respective eigenvalues and $Q_{b|y} $ and $R_{b'|y'}$ are projectors onto subspaces -- not necessarily one-dimensional -- of the local Hilbert space of Bob's system}. A sufficient condition for the observables to commute is $\left[ Q_{b|y},R_{b'|y'} \right] = 0$ for all $b$ and $b'$. We, thus, use this condition to define a pair of compatible projective measurements, as the following. Let $\left\{ Q_{b|y} \right\}$ and $\left\{ R_{b'|y'} \right\}$ be projective measurements labeled by $y$ and $y'$. They are compatible if, for all $b$ and $b'$, $\left[ Q_{b|y}, R_{b'|y'} \right] = 0$ holds. As previously, defining $\b = \left(b,b'\right)$ and $\y = \left(y,y'\right)$ allows us to write the joint projective measurement as $\left\{ S_{\b|\y} \right\}$, where each projector is given by $S_{\b|\y} = Q_{b|y}R_{b'|y'}$. This construction can be directly extended to contexts that involve more than two measurements.

The same definition can be extended to POVMs, as follows. Let $\left\{ Q_{b|y} \right\}$ and $\left\{ R_{b'|y'} \right\}$ be POVM measurements labeled by $y$ and $y'$, \ie, $Q_{b|y} \geq 0$ for all $b$, $R_{b'|y'} \geq 0$ for all $b'$, and $\sum_{b}Q_{b|y} = \sum_{b'}R_{b'|y'} = \mathds{1}$, the identity operator. Then, if, for all $b$ and $b'$, $\left[ Q_{b|y}, R_{b'|y'} \right] = 0$, we define measurements $y$ and $y'$ to be compatible, and a joint POVM can be defined as $\left\{ S_{\b|\y} \right\}$, where each POVM element is given by $S_{\b|\y} = Q_{b|y}R_{b'|y'}$ -- note that the commutation relations imply that the product of the positive semi-definite operators is itself positive semi-definite, and it is easy to see that $\sum_{\b} S_{\b|\y} = \sum_{b,b'} Q_{b|y}R_{b'|y'} = \mathds{1}$.

Now, consider a bipartite measurement scenario where one of the parties, Bob, is able to implement compatible measurements on his subsystem, as defined in Sec. \ref{sec:scenario} and depicted in the lower panel of Fig. \ref{fig:new_scenario}. Assuming the measurements are performed on a shared quantum system in state $\rho$, the joint probabilities of obtaining outcomes $a$ and $\b$ for respective measurements $x$ and $\y$ are given by
\begin{equation}
p(a,\b|x,\y) = \tr{\rho P_{a|x}\otimes S_{\b|\y}},
\end{equation}
where $\{P_{a|x}\}$ and $\{S_{\b|\y} \}$ are, in general, POVM measurements,  with, as previously discussed, the elements of the later given by $S_{\b|\y} = Q_{b|y}R_{b'|y'}$, where $\{Q_{b|y}\}$ and $\{R_{b'|y'}\}$ are POVMs associated to measurements $y$ and $y'$, respectively, respecting $\left[ Q_{b|y}, R_{b'|y'} \right] = 0$ for all $b$ and $b'$.

Regarding the nonlocality of quantum systems, an important question is whether scenarios with compatible measurements may activate the nonlocality of entangled states that are local in standard Bell scenarios. Let us first define locality of quantum states in standard Bell scenarios and in scenarios extended with compatibilities.

Let $\rho$ be a density operator acting on $\Hil = \Hil_{d_{A}} \otimes \Hil_{d_{B}}$, where $d_{A}$ and $d_{B}$ are the dimensions of the Hilbert spaces associated to the subsystems $A$ and $B$, respectively. We define $\rho$ to be \emph{local} if, \emph{for all} POVMs $\{P_{a|x}\}$ acting on $\Hil_{d_{A}}$ and $\{Q_{b|y}\}$ acting on $\Hil_{d_{B}}$, there exist a variable $\lambda$ in a set $\Lambda$ and probability distributions $q(\lambda)$, $p(a|x,\lambda)$ and $p(b|y,\lambda)$ such that
\begin{equation}\label{eq:rho_local}
\tr{\rho P_{a|x}\otimes Q_{b|y}} = \int_{\Lambda} p(a|x,\lambda) p(b|y,\lambda)q(\lambda)d\lambda.
\end{equation} 
If, in particular, Eq.~\eqref{eq:rho_local} holds for projective measurements, then we say $\rho$ is local with respect to projective measurements.

Now, let $\Hil_{d'_{B}}$ be a Hilbert space associated to subsystem $B$, where $d_{B}' \geq d_{B}$, and let $\rho'$ denote state $\rho$ trivially embedded in $\Hil = \Hil_{d_{A}}\otimes\Hil_{d'_{B}}$. We define $\rho$ to be \emph{local in an extended Bell scenario} -- or \emph{extended-local}, for short -- if, \emph{for all} POVMs $\{P_{a|x}\}$ acting on $\Hil_{d_{A}}$, and \emph{for all} compatible pairs of POVMs $\{Q_{b|y}\}$ and $\{R_{b'|y'}\}$ acting on $\Hil_{d_{B}'}$, there exist a variable $\lambda$ in a set $\Lambda$ and probability distributions $q(\lambda)$, $p(a|x,\lambda)$, and {no-disturbing probability distributions $p(b,b'|y,y',\lambda)$} such that
\begin{multline}
\tr{\rho' P_{a|x}\otimes Q_{b|y}R_{b'|y'}}  \\ = \int_{\Lambda} p(a|x,\lambda) p(b,b'|y,y',\lambda)q(\lambda)d\lambda.
\end{multline} 

{In Appendix \ref{app:locality} we prove the following:
{\newline \textbf{Theorem:} If $\rho$ is extended-local, then $\rho$ is local.}

Since the above theorem does not guarantee the equivalence between locality and extended-locality of quantum states, it could be the case that a local quantum state $\rho$ could lead to a nonlocal behavior in a scenario with compatible measurements; in other words, a scenario with compatible measurements could activate the nonlocality of $\rho$. However, even if the converse of the theorem is true and equivalence between locality and extended-locality as properties of quantum states holds, a scenario with compatible measurements could be more economical, in terms of the number of measurements, for instance, than a standard Bell scenario to display nonlocal behavior of a quantum state.

In the following section, we present, in some detail, an interesting and simple example of Bell scenario with local compatible measurements where these approaches and scenario provide an advantage for the detection of nonlocality of two important families of quantum states, as compared to similar standard Bell scenarios.}


\section{Application}
Consider a bipartite scenario where Alice is able to {choose between} two dichotomic measurements to perform, {so $\mathcal{A} = \{\pm1\}$,  $\mathcal{X} = \{0,1\}$, and Bob is able to perform four dichotomic measurements, $\mathcal{B} = \{\pm 1\}$ and $\mathcal{Y} = \{0,1,2,3\}$}, assumed to be compatible according to the contexts $\mathcal{C} = \{\{0,1\},\{1,2\},\{2,3\},\{3,0\}\}$. In this scenario, a behavior $\p$ will have $64$ components; however, due to normalization, no-signalling, and no-disturbance, only $26$ components will be independent. Due to this reason, it is convenient to work with correlators instead of probabilities. {Defining new random variables $A_{x}$, $B_{y}$ and $B_{\y} = B_{y_{1}}B_{y_{2}}$, valued on the set $\{\pm 1\}$, to represent the outcomes of the respective measurements}, the ``full'' correlators are defined as {the mean value of their product, as follows:}
{
\begin{equation}
\corr{A_{x}B_{\y}} = p(ab_{1}b_{2} = 1|x,\y) -  p(ab_{1}b_{2} = -1|x,\y),
\end{equation}
}
for all $x \in \mathcal{X}$ and $\y \in \mathcal{C}$, $(b_{1},b_{2})$ being the respective outcomes of measurements $(y_{1},y_{2}) = \y$. ``Marginal'' correlators $\corr{A_{x}}$, $\corr{B_{y}}$, $\corr{B_{\y}}$, $\corr{A_{x}B_{y}}$, for all $x \in \mathcal{X}$, $y \in \mathcal{Y}$ and $\y \in \mathcal{C}$, are analogously defined with the corresponding marginal probability distributions. The behaviors will, then, be vectors $\vec{c} \in \R^{d}$, where each component is a correlator. It is easy to check that there are exactly $26$ correlators in total, so $d = 26$; and, given the correlators, all the $64$ probabilities can be retrieved as
{
\begin{multline}
p(a,\b|x,\y) = \frac{1}{8}\left[1 + a \corr{A_{x}} + {b_{1}} \corr{B_{y_{1}}} + {b_{2}} \corr{B_{y_{2}}}  \right. \\ + {b_{1}b_{2}} \corr{B_{\y}}  + {ab_{1}} \corr{A_{x}B_{y_{1}}} \\ + \left. {ab_{2}} \corr{A_{x}B_{y_{2}}} + {ab_{1}b_{2}} \corr{A_{x}B_{\y}}\right].
\end{multline}
}

Now, we want to characterize the facets of the local and no-disturbance polytope of the scenario, {$\mathcal{P}_{AB}$}. The first step is to obtain all the extremal points of Bob's no-disturbance polytope. In all scenarios where the compatibility relations among dichotomic measurements are cyclic, it is known \cite{araujo2013} that the extremal points of the no-disturbance polytope, {up to outcome or measurement relabellings that respect the compatibility relations\footnote{The measurement relabellings that respect the compatibility relations are related to the group of isomorphisms of the {compatibility graph} of the scenario, a graph $G = (V,E)$ where each vertex $v_{i} \in V$ is associated to a measurement and each edge $ v_{i} \sim v_{k} = e \in E$ connects vertices associated to compatible measurements. The isomorphism group of cyclic graphs, such as the one considered in this section, is the dihedral group.},} are either of the form \footnote{Correlators that have this form are so-called \emph{noncontextual}, according to the literature on Kochen-Specker noncontextuality.}
\begin{subequations}\label{eq:ncpoint}
\begin{align}
&\corr{B_{y}} = \pm 1\\
& \corr{B_{\y}} = \prod_{y \in \y}\corr{B_y},
\end{align} 
\end{subequations}
for all $y \in \mathcal{Y}$ and $\y \in \mathcal{C}$, or {of the form}
\begin{subequations}\label{eq:prbox}
\begin{align}
& \corr{B_{y}} = 0, \, \forall \, y \in \mathcal{Y};  \\
& {\corr{B_{\y}} \in \{\pm 1 \},\, \forall \, \y \in \mathcal{C},\, \textnormal{s. t.}\,\prod_{\y}\corr{B_{\y}} = -1.}
\end{align}
\end{subequations} 
Then, the extremal points of the local, locally {no-disturbing} polytope will be behaviors whose bipartite correlators are of the form
\begin{subequations}
\begin{align}
& \corr{A_{x}B_{y}} = \corr{A_{x}}\corr{B_{y}}, \\ 
& \corr{A_{x}B_{\y}} = \corr{A_{x}}\corr{B_{\y}},
\end{align}
\end{subequations}
where $\corr{A_{x}} \in \{\pm 1\}$ and the behavior of Bob's box is given by either Eqs.~\eqref{eq:ncpoint} or Eqs.~\eqref{eq:prbox}, for all $x \in \mathcal{X}$, $y \in \mathcal{Y}$ and $\y \in \mathcal{C}$.

Having all the extremal points, we used \texttt{panda} to obtain the facets of the local, no-disturbance polytope. We found $26$ classes of inequalities, all of which are given in Appendix \ref{app:inequalities}. This result should be contrasted to the fact that, in standard bipartite Bell scenarios where no assumption regarding compatibility is made, if the number of measurements of one of the parties is $2$ and they are dichotomic, the only Bell inequality, up to rellabelings, is the CHSH inequality, as has been proven by Pironio in Ref.~\cite{Pironio2014}.

{Actually, the compatibility relations we assume can be implemented in a tripartite Bell scenario, if {we {assign} measurements $B_{0}$ and $B_{2}$ {to} one {party} {(say, {Bob$_0$})} and $B_{1}$ and $B_{3}$ {to} another {(say,  {Bob$_1$})}}. 
Due to this reason, some of the inequalities we obtain are equivalent to Sliwa's inequalities \cite{Sliwa} (see discussion in Appendix \ref{app:sliwa}), the Bell inequalities that completely characterize the local polytope in a tripartite scenario where each party is able to perform two dichotomic measurements. Note, however, that, had we assumed another compatibility structure for Bob's measurements, \eg, if the compatibility graph $G$ was a pentagon instead of a square, than it would not be possible to relate the scenario to any usual Bell scenario, {since it would not be possible to assign subsets of measurements to two or more parties in a way that is consistent with the the assumed compatibilities \footnote{{Measurement compatibility relations encoded in a compatibility graph $G$ correspond to a standard Bell scenario with $k$ space-like separated parties if and only if $G$ is a {complete k-partite graph}, defined as a graph $G = (V,E)$ where the set of vertices $V$ can be split into $k$ disjoint subsets $V_{i}$, such that $ v_{k} \sim v_{l} = e \in E$ if and only if $v_{k} \in V_{i}$ and $v_{l} \in V_{j \neq i}$.}}}.}

Among the $26$ inequalities we obtain, one has the form
\begin{equation}\label{eq:inequality}
2\corr{B_{0}} + \corr{(1-B_{0})[A_{0}(B_{1}+B_{3})+A_{1}(B_{1}-B_{3})]} \leq 2.
\end{equation}
Note that the term in square brackets corresponds to the left-hand side of a CHSH inequality between Alice and measurements $1$ and $3$ of Bob. To study the quantum violation of inequality Eq.~\eqref{eq:inequality}, it is convenient to define observables
{
\begin{subequations}
\begin{align}
A_{x} & = P_{+|x} - P_{-|x}, \\
B_{y} & = Q_{+|y} - Q_{-|y},
\end{align}
\end{subequations}
}
where $P_{a|x}$ and $Q_{b|y}$ are projectors associated to outcomes $a$ and $b$ of measurements $x$ and $y$, respectively, so the correlators will be evaluated as $\corr{A_{x}B_{\y}} = \tr{\rho A_{x}\otimes B_{\y}}$, where $B_{\y} = B_{y_{1}}B_{y_{2}}$, and $[B_{y_{1}},B_{y_{2}}] = 0$ for all $\y \in \mathcal{C}$. 

{Inequality Eq.~\eqref{eq:inequality} is equivalent to the class $\# 4$ of Sliwa \cite{Sliwa}. For quantum systems, it is maximally violated up to the value $4\sqrt{2}-2$, attained by a two-qubit maximally entangled state embedded in $\C^{2}\otimes\C^{4}$ \cite{sheila}. We now show that this inequality can certify the nonlocality of bipartite quantum states that do not violate the CHSH inequality.}

Consider the following two-parameter family of two-qubit states
\begin{subequations}\label{eq:horodecki_states1}
\begin{equation}
\rho\de{\alpha,w} = w\proj{\psi(\alpha)} + (1-w)\proj{00},
\end{equation}
where
\begin{equation}
\ket{\psi(\alpha)} = \sqrt{\alpha}\ket{01} + \sqrt{1-\alpha}\ket{10}.
\end{equation}
\end{subequations}
This family is known to include the two-qubit states with highest entanglement (as quantified by negativity and concurrence) that do not violate the CHSH inequality \cite{Bartkiewicz2013}. 
We, then, perform a seesaw optimization, embedding the states  in $\C^{2}\otimes\C^{4}$ to impose the compatibility relations among the measurements (details in the Appendix~\ref{app:seesaw}), and search for the {lowest value} of {$w$} such that the inequality is violated, for each $\alpha$. 
The results are displayed in Fig. \ref{fig:violation}, where we also plot the critical values of $w$ as a function of $\alpha$ for the CHSH inequality, provided by means of the Horodecki criterium \cite{Horodecki1995}, {and upper bounds on the critical values of $w$, obtained by means of a seesaw optimization, for the I$_{3322}$ inequality \cite{i3322} -- a relevant Bell inequality in the scenario where Alice and Bob perform three dichotomic measurements each -- , given by the expression
\begin{multline}
-\corr{A_{1}}-\corr{A_{2}}-\corr{B_{1}}-\corr{B_{2}}-\corr{A_{1}B_{1}}-\corr{A_{2}B_{1}}- \corr{A_{3}B_{1}} \\ - \corr{A_{1}B_{2}}-\corr{A_{2}B_{2}}+ \corr{A_{3}B_{2}}-\corr{A_{1}B_{3}}+\corr{A_{2}B_{3}} \leq 4.
\end{multline}
In fact, {the state $\rho\de{0.80,0.85}$ in family Eq.~\eqref{eq:horodecki_states1}  was the example considered in Ref.~\cite{i3322} of a state that does not violate the CHSH inequality, that, however, violates I$_{3322}$. 
In Ref.~\cite{liang2007bounds}, the authors show that, for $\alpha = 0.80$}, inequality I$_{3322}$ is violated for $w \gtrsim 0.837$, in excellent agreement with the value $0.838$ we obtain, corroborating with the precision of our lower bounds.}

\begin{figure}[htbp]
\includegraphics[height = 0.75\columnwidth,width = \columnwidth]{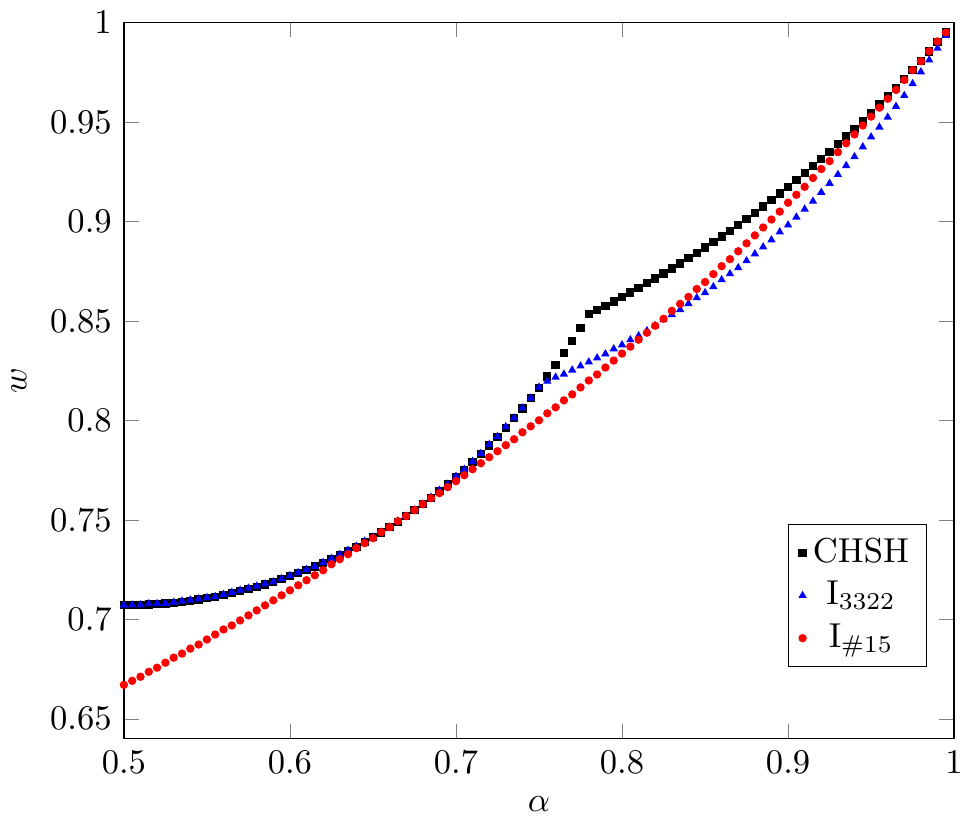}
\caption{{For states $\rho\de{\alpha,w}$ defined in Eq.~\eqref{eq:horodecki_states1}, the plot shows the critical parameter $w$, as a function of $\alpha$, above which inequality Eq.~\eqref{eq:inequality} is violated (red circles), above which the CHSH inequality is violated (black squares), {and above which the I$_{3322}$ inequality is violated (blue triangles)}. 
The points for inequalities Eq.~\eqref{eq:inequality} {and I$_{3322}$} were obtained via a see-saw optimization, and are, hence, upper bounds on the actual critical points. 
The points for the CHSH are exact, obtained by means of Horodecki's {necessary and sufficient criterium for violation of the CHSH inequality by two-qubit states} \cite{Horodecki1995}.
Inequality Eq.~\eqref{eq:inequality} is denoted $\textnormal{I}_{\# 15}$ for consistency with the appendices and the data related to the red points, which are available at Ref.~\cite{data}.}}
\label{fig:violation}
\end{figure}

{Now, consider the following two-parameter family of two-qubit states
\begin{subequations}\label{eq:mafalda_states1}
\begin{equation}
\sigma\de{\alpha,w} = w\proj{\psi(\alpha)} + (1-w)\mathds{1}/4,
\end{equation}
where, as previously,
\begin{equation}
\ket{\psi(\alpha)} = \sqrt{\alpha}\ket{01} + \sqrt{1-\alpha}\ket{10}.
\end{equation}
\end{subequations}
For $\alpha = 1/2$, the states obtained are locally equivalent to two-qubit Werner states, known to be entangled for $w > 1/3$, and local with respect to projective measurements for $w \lesssim 0.68$ \cite{Hirsch2017}. The best-known bounds on the locality of the states in family Eq.~\eqref{eq:mafalda_states1} are provided in Ref. \cite{Hirsch2016}. Applying the same methods adopted in the previous example, we were able to obtain upper bounds on the values of $w$, as a function of $\alpha$, above which states Eq.~\eqref{eq:mafalda_states1} violate inequality Eq.~\eqref{eq:inequality}. Results are shown in Fig. \ref{fig:violation2}, where it is clear that inequality Eq.~\eqref{eq:inequality} is better than the CHSH inequality to witness the nonlocality of this family of states specially in the range $0.7 < \alpha < 1$. Note that, compared to Fig. \ref{fig:violation}, points corresponding the the $I_{3322}$ inequality are absent, and this is due to the fact that $I_{3322}$ does not provide any advantage over the CHSH inequality to witness the nonlocality of this family of states.}

\begin{figure}[htbp]
\includegraphics[height = 0.75\columnwidth,width = \columnwidth]{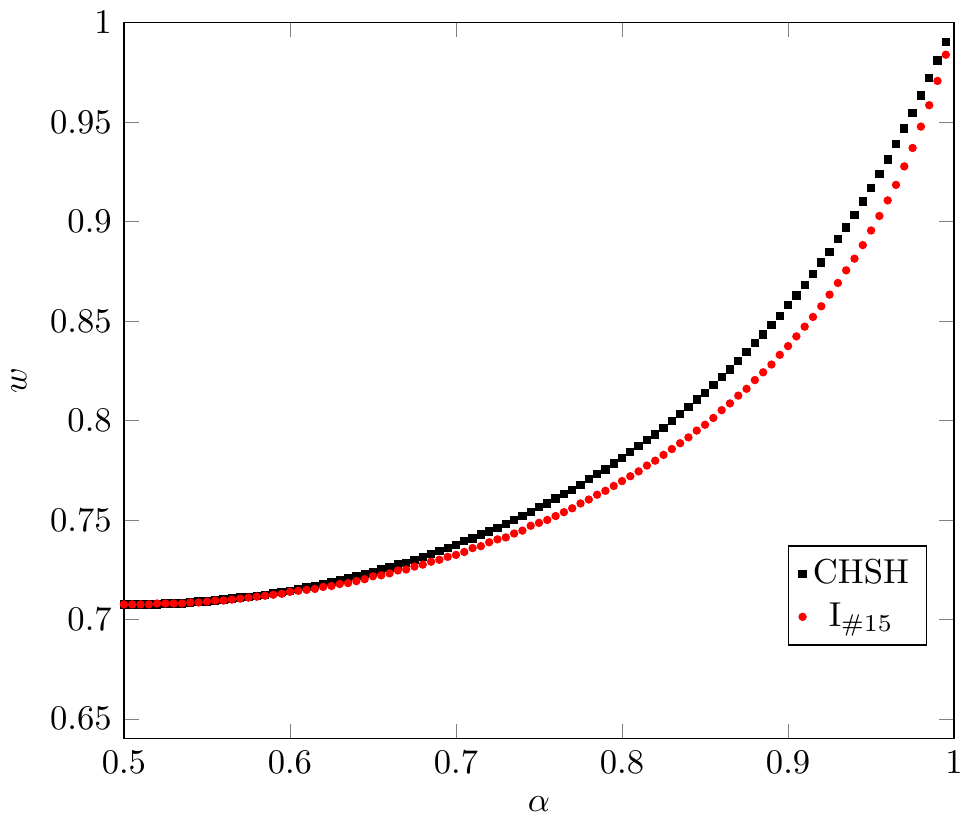}
\caption{{For states $\sigma\de{\alpha,w}$ defined in Eq.~\eqref{eq:mafalda_states1}, the plot shows the critical parameter $w$, as a function of $\alpha$, above which inequality \eqref{eq:inequality} is violated (red circles), and above which the CHSH inequality is violated (black squares). 
The points for inequality \eqref{eq:inequality} were obtained via a see-saw optmization, and are, hence, upper bounds on the actual critical points. 
The points for the CHSH are exact, obtained by means of Horodecki's {necessary and sufficient criterium for violation of the CHSH inequality by two-qubit states} \cite{Horodecki1995}.
As in Fig. \ref{fig:violation}, inequality \eqref{eq:inequality} is denoted $\textnormal{I}_{\# 15}$ for consistency with the appendices and the data related to the red points, which are available at \cite{data}.}}
\label{fig:violation2}
\end{figure}


\section{Discussion} 
{Although the simple examples we consider are sufficient evidence of the potential of the approach we introduce, it is only the first step in a direction for the study of Bell nonlocality. 
In principle, one could assume {a plethora of} more intricate local compatibility structures in scenarios with any number of parties, leading to a range of Bell-like inequalities.}

The {specific} compatibility structure we {consider} can be realized {-- with some loss of generality --} in a tripartite scenario, where each {party is able to perform two dichotomic measurements. An advantage of the tripartite implementation is that one does not need to \emph{assume} the compatibility relations; they would naturally hold due to space-like separation of the parties, implying that the test would be \emph{device-independent}.}
{Also, note} that {the locality assumption} in tripartite scenarios is more restrictive than the condition we demand {in our scenario}, since each $p\left( \b\middle| \y,\lambda\right)$ {is required} to obey the nondisturbance condition \eqref{eq:ND} {in the latter}, {as opposed to} strict locality, {in the former}.
This shows that our local and nondisturbance polytope is strictly larger than the corresponding tripartite local polytope; {more explicitly, notice that any vertex whose marginal behavior $p\left( \b\middle| \y\right)$ obeys Eqs.~\eqref{eq:prbox} is not tripartite-local}.

{One interesting fact, however, that is discussed in more detail in the Appendix \ref{app:sliwa}, is that some of the inequalities we obtain are isomorphic to the tripartite inequalities obtained by Sliwa \cite{Sliwa}, including Inequality Eq.~\eqref{eq:inequality}. This equivalence, together with the results presented in this manuscript, prove that there are multipartite Bell inequalities that are useful to witness the Bell nonlocality of bipartite quantum states {in a subtler way than just merging parts}.}

Two other scenarios that demand comparison are the ones obtained when we consider joint measurements of $B_y$ and $B_{y+1}$ (addition modulo $4$) as new measurements $B'_y$.
In the first case, if we consider outcomes {$b'_y = b_y b_{y+1}$}, Bob will have four mutually incompatible dichotomic measurements, and Ref. \cite{Pironio2014} shows that the only relevant inequalities for describing the local polytope belong to CHSH family.
Hence, our inequalities can show nonlocal behavior not revealed when such coarse grained version is considered.
The specific inequality Eq.~\eqref{eq:inequality}, for example, could never be written in such a scenario, since correlators like $\corr{B_y}$ or $\corr{A_xB_y}$ cannot be written as functions of the probabilities of the outcomes of $B'_y$.  
In the second case, if we consider $B''_y = \left(B_y , B_{y+1}\right)$, then Bob will have four mutually incompatible four-outcome measurements.
Once more, Ref. \cite{Pironio2014} implies that only CHSH inequalities are relevant for such a scenario, while the extra correlations coming from each $B_y$ being an element of $B''_y$ and $B''_{y-1}$ also would make it a somehow special realization of this Bell scenario (with such additional constraints).

{Also, it is worth mentioning that somewhat similar scenarios have been previously considered, with different focuses and assumptions. In Ref.~\cite{Gallego2014}, the authors present a formalism to study nonlocality in sequential measurement scenarios, mainly focused on the proper consequences due to the causal structures underlying the sequences of measurements. 
In Ref.~\cite{cabello2010proposal}, the {author argues, considering a bipartite scenario where one of the parties is able to perform sequential measurements, that local contextuality may lead to Bell nonlocality, although the definition of locality adopted is somehow intrincated }and seems to implicitly assume local noncontextuality.}


\section{Conclusion}
In this manuscript we present an approach to Bell nonlocality, an approach that takes into account the possibility that one (or more) of the parties is able to perform joint measurements according to given compatibility rules. 
We provide a precise definition of locality, or, more specifically, of local behaviors in these scenarios. Applying this definition, we completely characterize the set of local behaviors in the simplest scenario with compatible measurements, and we show how this approach leads to new, interesting Bell-like inequalities that may provide advantages over known Bell inequalities in witnessing the nonlocality of quantum states.
{We discuss in some detail two examples where such advantage appears; in particular, {both} families of states Eqs. ~\eqref{eq:horodecki_states1} {and \eqref{eq:mafalda_states1}} show nonlocal behavior in a scenario with compatible measurements for parameters where neither CHSH, nor $I_{3322}$, are able to witness it. {In the scenario considered, the compatibility relations can be implemented in a device-independent manner, and, thus, the examples show explicitly that this approach may provide advantages over standard Bell nonlocality for device-independent certification of entanglement.}}


\begin{acknowledgments}
R. R. thanks Jean-Daniel Bancal and Marcos C\'{e}sar de Oliveira. 
The authors thank the anonymous referees who contributed with suggestions that made the manuscript much better than its previous versions.
This work used computing resources and assistance of the John David Rogers Computing Center (CCJDR) in the Institute of Physics ``Gleb Wataghin'', University of Campinas.
{The authors acknowledge support from the Brazilian agencies CNPq and FAEPEX. R. R. is also supported by Grant No. 2018/07258-7, S\~{a}o Paulo Research Foundation (FAPESP).
This work is part of the Brazilian National Institute for Science and Technology on Quantum Information.}
\end{acknowledgments}



\begin{appendix}


\section{Locality of quantum states in scenarios with compatible measurements}\label{app:locality}

{In this Appendix, we prove that if a quantum state $\rho$ is local in a scenario with compatible measurements, it is local in a standard Bell scenario.} For simplicity, we have assumed a scenario where compatibilities are present only in party $B$, and each context is composed of two measurements. However, the following Theorem would still hold even under general definitions, which also take into account compatibilities in party $A$ and arbitrarily-sized contexts in both parties.

Before proceeding, let us introduce notation and important definitions. Let $\rho$ be a density operator acting on $\Hil = \Hil_{d_{A}} \otimes \Hil_{d_{B}}$, where $d_{A}$ and $d_{B}$ are the dimensions of the Hilbert spaces associated to the subsystems $A$ and $B$, respectively. We define $\rho$ to be \emph{local} if, \emph{for all} POVMs $\{P_{a|x}\}$ acting on $\Hil_{d_{A}}$ and $\{Q_{b|y}\}$ acting on $\Hil_{d_{B}}$, there exist a variable $\lambda$ in a set $\Lambda$ and probability distributions $q(\lambda)$, $p(a|x,\lambda)$ and $p(b|y,\lambda)$ such that
\begin{equation}
\tr{\rho P_{a|x}\otimes Q_{b|y}} = \int_{\Lambda} p(a|x,\lambda) p(b|y,\lambda)q(\lambda)d\lambda.
\end{equation} 
In the particular case where all measurements are assumed to be projective, we say $\rho$ is local with respect to projective measurements.

Now, let $\Hil_{d'_{B}}$ be a Hilbert space associated to subsystem $B$, where $d_{B}' \geq d_{B}$, and let $\rho'$ denote state $\rho$ trivially embedded in $\Hil = \Hil_{d_{A}}\otimes\Hil_{d'_{B}}$. We define $\rho$ to be \emph{local in an extended Bell scenario with compatible measurements in party $B$}, or \emph{extended-local}, for short, if, \emph{for all} POVMs $\{P_{a|x}\}$ acting on $\Hil_{d_{A}}$, and \emph{for all} compatible pairs of POVMs $\{Q_{b|y}\}$ and $\{R_{b'|y'}\}$ acting on $\Hil_{d_{B}'}$, there exist a variable $\lambda$ in a set $\Lambda$ and probability distributions $q(\lambda)$, $p(a|x,\lambda)$, and a no-disturbing probability distribution $p(b,b'|y,y',\lambda)$ such that
\begin{multline}\label{eq:ex_local}
\tr{\rho' P_{a|x}\otimes Q_{b|y}R_{b'|y'}} \\ = \int_{\Lambda} p(a|x,\lambda) p(b,b'|y,y',\lambda)q(\lambda)d\lambda.
\end{multline} 
In the particular case where all measurements are assumed to be projective, we say $\rho$ is extended-local with respect to projective measurements. 

{\textbf{Theorem:} If $\rho$ is extended-local, then $\rho$ is local.}

\textbf{Proof:} Let $\rho$ be extended-local, $d'_{B}= d_{B}$, and $R_{b'|y'} = \mathds{1}/n$, where $n$ denotes the number of outcomes of the measurement, for all $b'$ and $y'$, and $\mathds{1}$ denotes the identity operator in $\Hil_{d_{B}}$. By assumption, for all POVMs $\{Q_{b|y}\}$ acting on $\Hil_{d_{B}}$ Eq. \eqref{eq:ex_local} holds.
Now, marginalising over $b'$:
\begin{multline}
\sum_{b'}\tr{\rho P_{a|x}\otimes Q_{b|y}R_{b'|y'}}  \\ = \int_{\Lambda} p(a|x,\lambda) \left(\sum_{b'}p(b,b'|y,y',\lambda)\right)q(\lambda)d\lambda;
\end{multline}
leading to
\begin{equation}
\tr{\rho P_{a|x}\otimes Q_{b|y}} =  \int_{\Lambda} p(a|x,\lambda) p(b|y,\lambda)q(\lambda)d\lambda;
\end{equation}
which, by definition, holds for all POVMs $\{P_{a|x}\}$ acting on $\Hil_{d_{A}}$ and $\{Q_{b|y}\}$ acting on $\Hil_{d_{B}}$. Thus, $\rho$ is local.


\section{Inequalities}\label{app:inequalities}

In the scenario we have considered, Alice is able to perform two dichotomic measurements, and Bob is able to perform four dichotomic measurements; Bob's measurements, however, can be jointly performed according to compatibility rules provided by the contexts $\mathcal{C} = \{\{0,1\},\{1,2\},\{2,3\},\{3,0\}\}$.

Using the representation of correlators, we have listed all extremal points of the polytope of behaviors that are local, according to the definition we provided in the main text, such that Bob's marginal behaviors respect the no-disturbance conditions. 
Bob's {extremal} marginal behaviors belong to either one of two distinct classes:
\begin{itemize}
\item[(i)] noncontextual behaviors: are behaviors of the form
\begin{subequations}\label{appeq:ncpoint}
\begin{align}
&\corr{B_{y}} = \pm 1\\
& \corr{B_{\y}} = \prod_{{y \in \y}}\corr{B_y},
\end{align} 
\end{subequations}
for all $y \in \mathcal{Y}$ and $\y \in \mathcal{C}$;
\item[(ii)] contextual, no-disturbing behaviors: are behaviors of the form 
\begin{subequations}\label{appeq:prbox}
\begin{align}
& \corr{B_{y}} = 0, \, \forall \, y \in \mathcal{Y};  \\
& {\corr{B_{\y}} \in \{\pm 1 \},\, \forall \, \y \in \mathcal{C},\, \textnormal{s. t.}\,\prod_{\y}\corr{B_{\y}} = -1.}
\label{appeq:prbox-corr}
\end{align}
\end{subequations}
\end{itemize}
Then, the extremal points of the local, locally {no-disturbing} polytope will be behaviors whose bipartite correlators are of the form
\begin{subequations}
\begin{align}
& \corr{A_{x}B_{y}} = \corr{A_{x}}\corr{B_{y}}, \\ 
& \corr{A_{x}B_{\y}} = \corr{A_{x}}\corr{B_{\y}},
\end{align}
\end{subequations}
where $\corr{A_{x}} \in \{\pm 1\}$ and the behavior of Bob's box is given by either Eqs.~\eqref{appeq:ncpoint} or Eqs.~\eqref{appeq:prbox}, for all $x \in \mathcal{X}$, $y \in \mathcal{Y}$ and $\y \in \mathcal{C}$.

Having listed all the extremal points of the the local, no-disturbance polytope, we used the software \texttt{panda} \cite{panda} to change the representation of the polytope, and we obtained $26$ inequalities, up to rellabelings that respect the local compatibility rules. These inequalities are listed in Table \ref{tab:inequalities}.
This table should be read as follows: each row represents an inequality, labeled by the number in the first column. Each column, then, has the coefficient of the correlator represented in the heading, where the measurements $x$, $y_{1}$, and $y_{2}$ are the corresponding numbers in the second row. The second to last column refers to the local bound $\beta_{L}$ of each inequality.
In the last column, we list the quantum maxima $\beta_{Q}$ (exact up to the given precision) of each inequality. The maxima were upper bounded by means of the Navascu\'{e}s-Pironio-Ac\'{i}n \cite{NPA} hierarchy of semidefinite programs that outerapproximate the set of quantum correlations, implemented in \texttt{python} with the aid of the \texttt{NCPOL2SDPA} \cite{ncpol2sdpa} library. The values listed correspond to the third level of the hierarchy, and the optimizations were performed with the \texttt{MOSEK} \cite{mosek} solver. We have also computed lower bounds on the maxima by means of a seesaw optimization, detailed in Appendix \ref{app:seesaw}. 
The lower and upper bounds on quantum maxima obtained differ by less than $5 \times 10^{-4}$ for all inequalities; on average, they differ by $3 \times 10^{-5}$. For the 15 inequalities that are equivalent to Sliwa's inequalities, our results are in perfect agreement with those of \cite{sheila, vallins}, where, among other results, quantum maxima are computed and analyzed for all Sliwa's inequalities.

As an example, consider inequality 25. It is
\begin{multline}
\corr{A_{0}B_{0}}  + \corr{A_{0}B_{2}} + \corr{A_{1}B_{0}} + \corr{A_{1}B_{2}} + 2\corr{A_{0}B_{0}B_{1}} \\ + \corr{A_{0}B_{2}B_{3}} - \corr{A_{0}B_{3}B_{0}} - 2\corr{A_{1}B_{0}B_{1}} \\ + \corr{A_{1}B_{2}B_{3}} - \corr{A_{1}B_{3}B_{0}} \leq_{_{L}} 4 \leq_{_{Q}} 5.6568
\end{multline}

\section{Relation to Sliwa's inequalities}\label{app:sliwa}
Note that the particular scenario we consider {share some similarity with} a tripartite Bell scenario, where each party is able to perform two dichotomic measurements; the correspondence becomes  {explicit if one considers $B_0$ and $B_2$ as the possible measurements of a second {party}, while $B_{1}$ and $B_{3}$ are the possible choices of} a third party. 
This Bell scenario has been studied by Sliwa \cite{Sliwa}, who obtained 46 distinct classes of Bell inequalities by assuming full locality between the three parties. Had we considered only noncontextual marginal behaviors of Bob, Eq.~\eqref{eq:ncpoint}, the inequalities obtained would be all equivalent to Sliwa's 46 inequalities. 
However, by including points of the form of Eq.~\eqref{eq:prbox} (i) we obtain inequalities that are not equivalent to Sliwa's; and (ii) any violation of the inequalities is a certification of {bipartite} nonlocality, a fact that would not be true otherwise, since stronger-than-classical (contextual) correlations in the marginal behavior of Bob could lead to violations of the inequalities obtained solely via Eq.~\eqref{eq:ncpoint}.

Sliwa's inequalities are, as discussed, related to a polytope that is contained in the local, no-disturbance polytope we characterize. It would not be surprising, then, if some of the facets of the polytope were equivalent to the inequalities of Sliwa, and this is exactly what we observe. For the 15 inequalities we obtained that are equivalent to an inequality of Sliwa, we provide in the second column of Table I the number referring to the enumeration in Ref.~\cite{Sliwa}.

{
\begin{table*}
\begin{center}
\begin{ruledtabular}
\begin{tabular}{|c|c|cc|cccc|cccccccc|cccc|cccccccc|c|c|}
$\#$ & \# S & \multicolumn{2}{C|}{\corr{A_{x}}} & \multicolumn{4}{C|}{\corr{B_{y}}} & \multicolumn{8}{C|}{\corr{A_{x}B_{y}}} & \multicolumn{4}{C|}{\corr{B_{y_{1}}B_{y_{2}}}} & \multicolumn{8}{C|}{\corr{A_{x}B_{y_{1}}B_{y_{2}}}}& $\beta_{L}$ & $\beta_{Q}$ \\
& & 0 & 1 & 0 & 1 & 2 & 3 & 00 & 01 & 02 & 03 & 10 & 11 & 12 & 13 & 01 & 12 & 23 & 30 & 001 & 012 & 023 & 030 & 101 & 112 & 123 & 130 & &  \\\hline 1&1&1&0&1&1&0&0&-1&-1&0&0&0&0&0&0&-1&0&0&0&1&0&0&0&0&0&0&0&1&1.000\\
2&-&2&0&1&0&1&0&0&2&0&0&1&0&-1&0&0&0&1&1&-1&-1&-1&-1&-1&1&0&0&4&5.656\\
3&-&2&0&1&0&1&0&0&0&0&0&1&2&-1&0&0&0&1&1&-1&1&-1&-1&-1&-1&0&0&4&5.000\\
4&-&2&0&0&0&0&0&2&1&0&1&0&1&0&-1&-1&1&1&-1&0&-1&-1&0&-1&0&0&1&4&5.656\\
5&-&2&0&0&0&0&0&1&1&1&1&1&1&-1&-1&-1&1&-1&1&0&-1&0&-1&-1&0&1&0&4&5.753\\
6&-&2&0&0&0&0&0&0&0&0&0&2&2&0&0&-1&1&1&-1&-1&-1&-1&1&0&0&0&0&4&5.000\\
7&-&1&1&1&0&1&0&1&1&0&0&0&-1&-1&2&-1&-1&0&0&0&1&1&-1&1&0&-1&-1&4&5.656\\
8&-&1&1&1&0&1&0&1&2&0&1&0&0&-1&-1&0&0&-1&-1&-1&-1&1&0&-1&1&0&1&4&5.753\\
9&-&1&1&0&0&0&0&2&2&1&1&0&0&-1&-1&-1&-1&-1&-1&0&0&1&0&-1&1&0&1&4&5.656\\
10&-&1&1&0&0&0&0&2&1&1&0&0&1&-1&0&-1&1&1&-1&0&-1&0&0&-1&0&-1&1&4&5.656\\
11&17&1&1&0&0&0&0&1&1&0&0&1&1&0&0&0&0&0&0&-1&0&2&0&-1&0&-2&0&4&5.656\\
12&6&1&0&1&1&0&0&0&0&1&1&1&-1&-1&1&1&0&0&0&-1&-1&0&-1&0&1&0&-1&3&4.656\\
13&-&1&0&1&1&0&0&0&0&1&1&1&1&1&1&-1&0&0&0&1&-1&-2&-1&0&-1&2&-1&5&7.012\\
14&-&1&0&1&1&0&0&0&0&1&1&1&1&1&1&-1&0&0&0&1&-1&2&-1&0&-1&-2&-1&5&6.656\\
15&4&0&0&2&0&0&0&0&1&0&1&0&1&0&-1&0&0&0&0&-1&0&0&-1&-1&0&0&1&2&3.656\\
16&19&0&0&1&0&1&0&1&2&1&2&0&0&0&0&0&0&-1&-1&-1&-1&0&0&1&-1&1&-1&4&5.782\\
17&18&0&0&1&0&1&0&1&2&1&0&0&0&0&2&-1&-1&0&0&0&0&1&-1&1&-1&-1&-1&4&5.753\\
18&15&0&0&0&0&0&0&2&2&2&2&0&0&0&0&-1&-1&-1&-1&0&0&0&0&1&-1&1&-1&4&6.000\\
19&14&0&0&0&0&0&0&2&1&0&1&2&-1&0&-1&0&0&0&0&0&1&-1&0&0&-1&1&0&4&5.656\\
20&12&0&0&0&0&0&0&2&1&0&1&0&1&2&1&-1&-1&-1&-1&0&1&-1&0&-1&0&0&1&4&5.656\\
21&14&0&0&0&0&0&0&2&0&2&0&0&0&0&0&1&-1&1&-1&0&0&0&0&1&-1&-1&1&4&5.656\\
22&13&0&0&0&0&0&0&2&0&2&0&0&0&0&0&0&0&0&0&1&-1&1&-1&1&-1&-1&1&4&5.656\\
23&11&0&0&0&0&0&0&2&0&0&0&0&0&2&0&0&0&0&0&1&1&1&-1&-1&-1&1&-1&4&5.656\\
24&10&0&0&0&0&0&0&1&1&1&1&1&-1&1&-1&1&-1&1&-1&1&0&-1&0&0&-1&0&1&4&4.000\\
25&9&0&0&0&0&0&0&1&0&1&0&1&0&1&0&0&0&0&0&2&0&1&-1&-2&0&1&-1&4&5.656\\
26&3&0&0&0&0&0&0&0&0&0&0&0&0&0&0&0&0&0&0&1&0&1&0&1&0&-1&0&2&2.828
\end{tabular}
\caption{All 26 classes of inequalities that are facets of the local, no-disturbing polytope of the measurement scenario we introduce, displayed as coefficients of correlators. Some of the inequalities are equivalent to Sliwa's inequalities \cite{Sliwa}; the corresponding class in Ref.~\cite{Sliwa} is displayed in the second column, $\# S$. The second-to-last column displays the local bounds of the inequalities, and the last column displays their respective maximal quantum violations, exact, up to the given precision.}\label{tab:inequalities}
\end{ruledtabular}
\end{center}
\end{table*}}


\section{Seesaw optimization}\label{app:seesaw}

To compute lower bounds on the maximum quantum violation of the 26 inequalities we study, as well as the bounds on the critical parameters of the family of quantum states we present in the main text, we implemented variations of an optmization algorithm known as see-saw iteration, introduced by Werner and Wolf in Ref.~\cite{WernerWolf}. Our implementation follows the steps described in Sec. II.B.3 of Ref.~\cite{liang2007bounds}, with minor adjustments.

For standard Bell inequalities where the parties perform dichotomic measurements, the algorithm is based on the idea that, if the quantum state and the measurements of all but one of the parties are fixed, then optmization over the measurements of the remaining party can be carried out explicitly. Consider, for clarity, a bipartite scenario; extensions to multipartite ones are straightforward. Let the operator associated to a given Bell inequality be
\begin{equation}
\beta = \sum_{x=1}^{m_{A}} \sum_{y=1}^{m_{B}} {\sum_{a=-1}^{1} \sum_{b=-1}^{1}}c_{x,y}^{a,b} P_{a|x} \otimes Q_{b|y}, 
\end{equation}
where $x$ ($y$) labels the choice among the $m_{A}$ ($m_{B}$) possible measurements of party $A$ ($B$), $a$ ($b$) labels the possible outcomes, $P_{a|x}$ ($Q_{b|y}$) is the measurement operator associated to outcome $a$ ($b$) of measurement $x$ ($y$), and $c_{x,y}^{a,b}$ are the respective coefficients that define the inequality. Then, the quantum average value of the inequality can be written, as a function of the state $\rho$ and the measurement operators, as
\begin{subequations}
\begin{equation}
S_{P}(\rho,\DE{P_{a|x}},\DE{Q_{b|y}}) = \sum_{b,y}\trd{\rho_{Q_{b|y}}Q_{b|y}},
\end{equation}
where
\begin{equation}
\rho_{Q_{b|y}} = \sum_{a,x} c_{x,y}^{a,b} \trad{\rho(P_{a|x}\otimes \id)},
\end{equation}
\end{subequations}
where $\tra{.}$ denotes the partial trace over subsystem $A$. For fixed $\rho$ and $P_{a|x}$, $S_{P}$ is a linear function of $Q_{b|y}$. And, since $Q_{1,y} = \id - Q_{-1|y}$, we have
\begin{multline}
\sum_{b = \pm 1} \tr{\rho_{Q_{b|y}}Q_{b|y}} = \\ \tr{\left( \rho_{Q_{+1|y}} - \rho_{Q_{-1|y}} \right) Q_{+1|y}} + \tr{\rho_{Q_{-1|y}}}.
\end{multline}
This expression can be optimized by setting $Q_{+1|y}$ equal to the projector onto the positive subspace of $\rho_{Q_{+1|y}} - \rho_{Q_{-1|y}}$. This procedure can, then, be iterated, so optimization can be carried over all measurements of all parties. If desired, then the quantum state can be optimized over, in an even simpler fashion: In any step, the optimal quantum state {can be taken as a pure state given by an} eigenvector of $\beta$ associated to its maximal eigenvalue.

Note that the first step of the seesaw algorithm already requires {a choice of} state and measurements, so they should be randomly generated in the beginning of the process. 
Although it is clear that the algorithm will converge after a sufficient number of steps, one cannot guarantee that it will converge to the global maximum of the problem. 
Any solution, however, is a lower bound to the optimal solution, so it is recommended to restart the algorithm with as many random ``seeds'' as feasible.

The scenario we consider, as discussed in the previous section, is similar to a tripartite scenario where each of the three parties is able to perform two dichotomic measurements. Our implementation makes use of this similarity, assuming measurements $B_{1}$ and $B_{3}$ are implemented by a third party. On one hand, this assumption guarantees the compatibility relations assumed in the scenario; on the other hand, it leads to loss of generality. This is one more reason (despite the fact that the see-saw does not necessarily converge to the global maximum) that advocates against optimality of the bounds computed via this method.

Our goal was to compute upper bounds on the critical values of $w$, as a function of $\alpha$, such that the {two families of} states
\begin{subequations}
\begin{equation}\label{eq:horodecki_states}
\rho = w\proj{\psi(\alpha)} + (1-w)\proj{00},
\end{equation}
and 
\begin{equation}\label{eq:mafalda_states}
\sigma\de{\alpha,w} = w\proj{\psi(\alpha)} + (1-w)\mathds{1}/4,
\end{equation}
where
\begin{equation}
\ket{\psi(\alpha)} = \sqrt{\alpha}\ket{01} + \sqrt{1-\alpha}\ket{10},
\end{equation}
\end{subequations}
violates inequality $\# 15$ (we have numerical evidence that this is the best inequality among the ones we listed to witness the nonlocality of such states). The code was implemented in \texttt{MATLAB}, with the aid of the \texttt{QETLAB} \cite{qetlab} library. {In both cases,} we suppose a system with local Hilbert spaces $\mathcal{H}_{A} = \C^2$ and ${\mathcal{H}_B = \C^4}$, where the states Eqs.~\eqref{eq:horodecki_states}  and {\eqref{eq:mafalda_states}} are embedded trivially {(meaning that, for party $B$, the elements of the computational basis of $\C^2$, according to which states Eqs.~\eqref{eq:horodecki_states} and \eqref{eq:mafalda_states} are defined, are mapped to two elements of the computational basis of $\C^4$)}, with $B_{i} = \tilde{B}_{i}\otimes \id_{2}$, for $i \in \{0,2\}$, and $B_{j} = \id_{2}\otimes \tilde{B}_{j}$, for $j \in \{1,3\}$, where $\tilde{B}_{i}$ acts in $\C^2$ and $\id_{2}$ is the identity operator in the same space. Then, for each of $100$ values of $\alpha$ equally spaced in the interval $[1/2,1]$, we start with $w=3/4$ and run the seesaw with at most $500$ random `seeds' -- projective measurements for all parties, and a {random} local unitary $U$ acting on $\C^4$ that we apply to the state, so it is not always fixed in the same basis as the virtual parties $B$ and $C$ are divided. The process is iterated eight times for different values of $w$, which is updated according to a bissection scheme: If a violation of the inequality is obtained in iteration $i$, then value of $w$ is updated to ${w-2^{-(i+2)}}$; if, after all seeds, no violation is obtained, then the value of $w$ is updated to ${w+2^{-(i+2)}}$.
\end{appendix}


\end{document}